\documentclass[pss]{wiley2sp}
\usepackage{amsmath}

\begin{document}
\titlefigure[width=\linewidth,height=3.1cm]{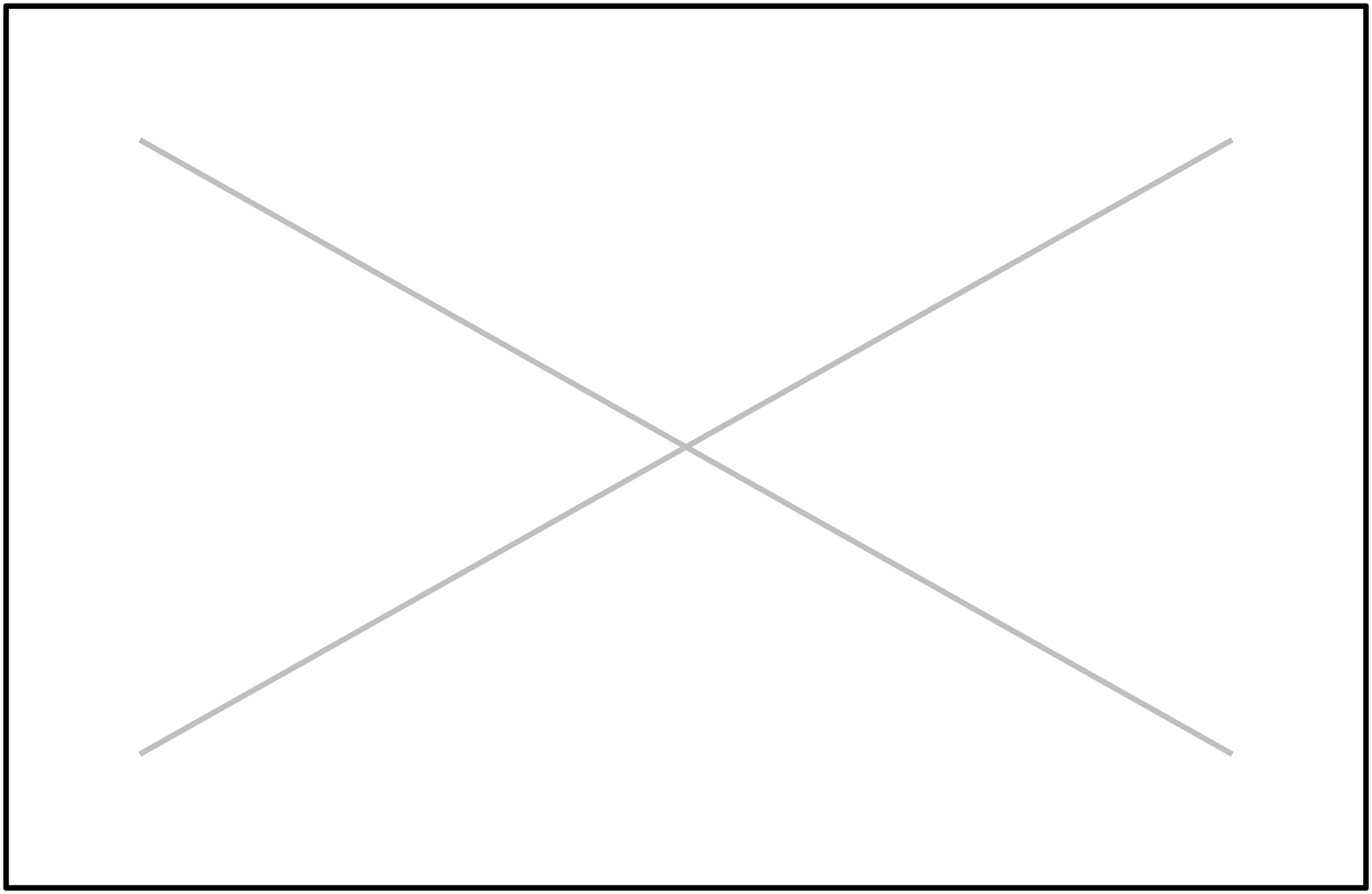}
\titlefigurecaption{This is the figure caption.}
\title{Structural peculiarities of plastically-deformed cementite and their
influence on magnetic characteristics and Mossbauer parameters}
\titlerunning{Structural peculiarities of plastically-deformed cementite}
\author{%
  A.K. Arzhnikov\textsuperscript{\Ast}, 
  L.V. Dobysheva}
\authorrunning{A.K. Arzhnikov et al.}
%\def\mailname{Standard is "Corresponding author"}
%E-mail-address of corresponding author
\mail{e-mail
  \textsf{arzhnikov@otf.pti.udm.ru}, Phone
  +07-3412-218988, Fax +07-3412-250614}
\institute{%
  Physical-Technical Institute, 
Ural Branch of Russian Academy of Sciences, 
Kirov str.~132, Izhevsk 426001, Russia}
\received{XXXX, revised XXXX, accepted XXXX}
\published{XXXX}
\abstract{% 10 lines !!!
Cementite Fe3C is studied with first-principles calculations. Two
possible positions of carbon atoms in the iron sublattice are
considered: with prismatic or octahedral environment. Mossbauer spectra
(MS) with parameters calculated for both modifications are simulated
above and below the Curie temperature. A possibility to detect the
change in carbon position upon annealing from MS is discussed. It is
shown that this is hardly possible using a standard approach to 
treatment of MS, but it can be seen in more subtle details of the MS
below the Curie temperature, such as widths and positions of separate
lines.}
\maketitle
% \Dsection{First section}
%%% 1 %%%
%\section{Introduction} 
The properties of cementite, i.e. iron carbide $Fe_3C$, are being
studied for a long time. Originally, cementite was investigated in order
to improve the mechanical properties of steels. Its magnetic properties
have received some attention in connection with nondestructive methods
of the steel control. A recent interest in the $Fe_3C$ properties is
aroused by new possibilities of obtaining metastable compounds by
mechanical alloying, implantation and so forth (see, for example, 
\cite{Els1,Reed,Koniger,Umemotoa}). An additional impact has been given
by the studies dealing with the chemical composition of the Earth's core
\cite{Wood1,Wood2}. Besides, the monophase $Fe_3C$ by itself has
attracted considerable interest due to some peculiarities in the bulk-modulus
behaviour \cite{Duman1} and the instability of the magnetic state under
pressure \cite{Lin,Duman2}.
\begin{figure}[!ht] 
\includegraphics*[width=8.8cm]{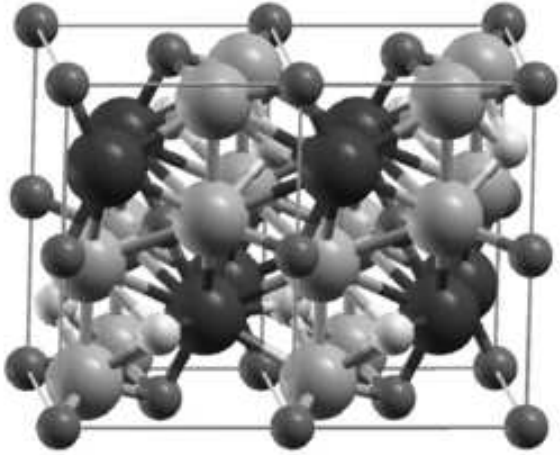}
\caption{Double unit cell of cementite. Large black and grey balls show
iron FeI (S  position) and FeII (G position) atoms; small white or dark
grey  balls represent carbon in prismatic or octahedral environment,
correspondingly.}
\label{cell-prism-oct}\end{figure}

The structure of cementite is believed to be known well from the
experiments on the X-ray diffraction, for example, \cite{Fasiska}. But
many magnetic, Mossbauer and EELS experimental data
\cite{Els2,Maratk,Els3}, conducted for non-equilibrium samples, lead us
to think that its structure varies depending on the mechanical and
thermal treatment. As these variations are not detected by the X-ray
diffraction, one can suppose that there are cementite modifications
differing in the carbon positions, which do not noticeably affect the
X-ray diffraction patterns because of the large difference in the
scattering factors between iron and carbon. The neutron diffraction
experiments conducted recently on the well-annealed samples
\cite{Wood-neutr} showed that carbon atoms occupy the prismatic
positions. So, after annealing at high temperatures (above 700 K
\cite{Fasiska,Wood-neutr,Els2}) cementite is in the ground state and has
a sole, well-determined, structure. The structure of the non-equilibrium
cementite comes into question.

According to the structural considerations in \cite{Schastl2}, carbon
can occupy four positions between the iron sites, that are called
prismatic and octahedral, and distorted prismatic and octahedral. The
carbon in the two last positions is closer to iron atoms and these
modifications appear to be improbable. Fig.~\ref{cell-prism-oct}
displays the structure of cementite, with two possible carbon positions
shown together.  

In this work, first-principles calculations have been conducted for
cementite with different positions of carbon in the lattice. Earlier  in
\cite{Our-CM}, using such calculations, we have received that the
prismatic modification of cementite has a smallest energy and is its
ground state, and explained an unusual behavior of the coercive force 
in the mechanically alloyed cementite on the annealing temperature as a
transition of cementite from octahedral modification with small
magneto-crystalline energy $E_{\rm MA}$ to the prismatic one with large
$E_{\rm MA}$. At the same time, we do not know unambiguous direct
experiments demonstrating existence of the octahedral modification of
cementite. The X-ray diffraction experiments do not show the carbon 
position with a necessary accuracy, and the neutron powder diffraction 
experiments have been done for well-annealed samples being in the ground 
state without traces of metastable phases. So, the aim of the paper is
to calculate and compare the parameters of hyperfine interaction of
cementite with different position of carbon atoms, to simulate the
Mossbauer spectra (MS) and to estimate the possibility of experimental
detection of the octahedral modification of cementite with the help of
the Mossbauer spectroscopy.

Some theoretical investigations of cementite have been already carried
out (see, for example, \cite{Wood2,Lin,Medvedeva,Haglund}). Only in
\cite{Medvedeva} an attempt to obtain and compare some characteristics
of the prismatic and octahedral cementite modifications has been done,
but no analysis on experimental detection has been conducted.

The crystal structure of $Fe_3C$ is an orthorhombic lattice, we use the
Pbnm space group. The unit cell contains 12 iron atoms and 4 carbon
atoms (a double unit cell, drawn with the help of the package 
\cite{xcrysden} in Fig.~\ref{cell-prism-oct}, shows carbon positions in
both prismatic and octahedral environment). 

The calculations presented in this paper have been conducted by the
full-potential linearized augmented plane wave (FLAPW) method in the
WIEN2k package \cite{WIEN2k}. The generalized gradient approximation of
the  exchange-correlation potential (GGA) \cite{GGA} is used in this
work. The details of the calculational procedure can be seen in
\cite{Our-CM}. Calculated are the Fe atomic magnetic moments, hyperfine 
magnetic fields (HFF), isomer shifts (IS), the electric field gradient 
(EFG) at the Fe nuclei. $V_{\rm zz}$ and  $V_{\rm xx}$ are the maximum-
and the minimum-in-magnitude components of the tensor of the EFG; $\eta
= (V_{\rm xx}-V_{\rm yy})/V_{\rm zz}$; $\theta$ and $\phi$ are the
polar and azimuthal angles of the HFF in the local coordinate system of
the EFG. Using the parameters calculated, we have simulated the
Mossbauer spectra for polycrystals (which is usually characteristic of
cementite) with the help of the program described in \cite{Kuz'min}.

Table~\ref{calc-HFparam} gives data (cell parameters, magnetic moments,
Mossbauer parameters) for each system. 
%\begin{table*}[width=\textwidth]
\begin{table}[b]
\caption{\label{calc-HFparam} The calculated structural parameters of the 
unit cells - lattice parameters a, b and c (nm) and  positions
(fractional  coordinates) of the atoms, spin $M_{spin}$ and orbital
$M_{orb}$  magnetic moments ($\mu_B$), the Fermi-contact $H_{spin}$ and
orbital  $H_{orb}$ contributions to the hyperfine magnetic field 
$H_{tot}=H_{spin}+H_{orb}$ (T), isomer shift IS ($mm/s$), electric field
gradient EFG $V_{zz}$ ($V/m^2$),  asymmetry parameter $\eta$, polar and
azimuthal angle between $H_{tot}$  and main axis of the EFG tensor.}
\small
\begin{tabular}{|c|c|c|c|c|}
\hline
&\multicolumn{2}{c|}{prismatic}&\multicolumn{2}{c|}{octahedral} \\
\hline
a, b, c &\multicolumn{2}{c|}{0.4490, 0.5047, 0.6743} 
                      &\multicolumn{2}{c|}{0.4711, 0.5111, 0.6960} \\
\hline
number &Fe I, & Fe II, & Fe I, & Fe II, \\
of atoms&
 4 atoms&  8 atoms& 4 atoms& 8 atoms  \\
\hline
$M_{spin}$& 1.97& 1.91& 2.09& 1.90  \\
$M_{orb}$ & 0.05& 0.04& 0.05& 0.03 \\
\hline
$H_{spin}$& -25.1& -24.7& -24.3& -23.7 \\
$H_{orb}$ &   3.1&   2.7&   3.1&   1.9 \\
$H_{tot}$ & -22.0& -22.0& -21.2& -21.8 \\
\hline
IS        & 0.16 & 0.17 & 0.22 & 0.14 \\
\hline
$V_{zz}$  & 3.07 & 1.37 & 1.99 &-2.09 \\
$\eta$    & 0.05 & 0.83 & 0.996& 0.23 \\
$\phi$    &  90  &138.6 &  90  &  73.1 \\
$\theta$  &   0  & 11.2 &  90  &  26.6 \\
\hline
\end{tabular}
     \end{table}
A difference between the magnetic moments in the prismatic and
octahedral cementite is found numerically small, so magnetic
measurements are not expected to determine which carbon position is
realized in samples. 

\begin{figure*}[!ht]
\sidecaption
\includegraphics*[width=.76\textwidth]{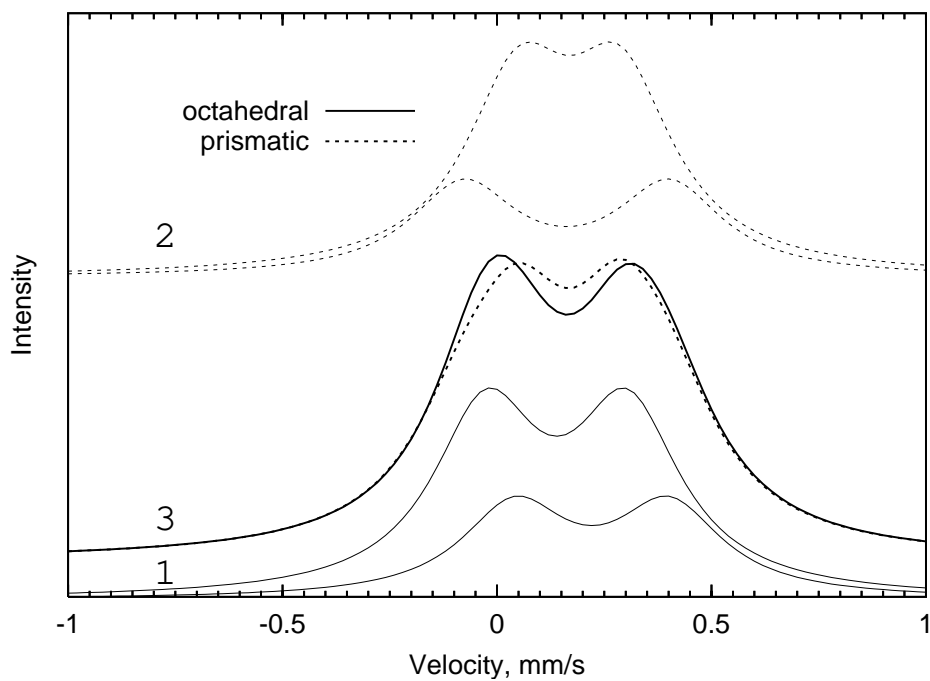}\hfill
\caption{The simulated Mossbauer spectra of cementite with parameters
from Table 1 above the Curie temperature (the HFF's equal zero). Curves 1
are the subspectra of two positions of iron in the octahedral 
modification, curves 2 are for the prismatic modification, curves 3  are
for the summary spectra of the octahedral or prismatic modifications.}
\label{Moss_aTc}\end{figure*} 
\begin{figure*}[!ht] 
\sidecaption
\includegraphics*[width=.76\textwidth]{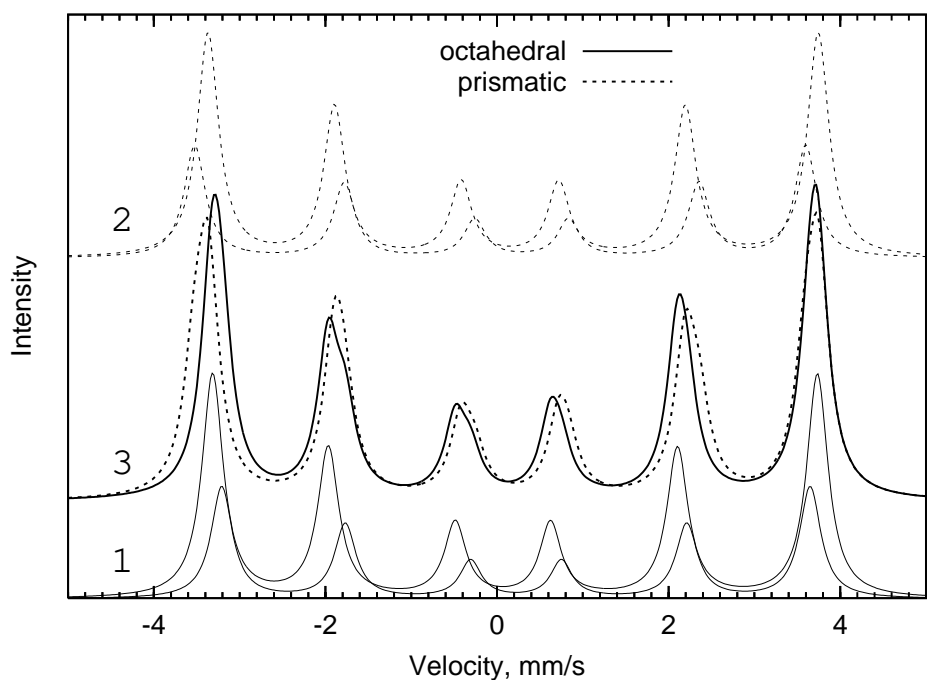}
\caption{The simulated Mossbauer spectra below the Curie temperature. 
Designations are as in Fig.2}
\label{Moss_bTc}\end{figure*}

A difference in the EFG between the two modifications draws attention.
One could expect that this difference should manifest itself in the
MS above the Curie temperature. The simulation of the
spectra with zero hyperfine field and other parameters from Table 1 (see
Fig.~\ref{Moss_aTc}) shows, however, that practically no difference can
be seen between the two modifications. This occurs because the
difference wears off due to summation of two subspectra from 
nonequivalent positions Fe I and Fe II.

Below the Curie temperature (see Fig.~\ref{Moss_bTc}), the spectra can
be distinguished  better, as the difference becomes clearer in the
outermost lines. The subspectra are essentially affected by the
combined hyperfine magnetic and quadrupole interaction. The summary
spectra cannot be also decomposed at a glance, but some traces of a
combined interaction still exist and manifest themselves in the
distances between the peaks of the simulated MS (see
Table~\ref{spectra-param}). We remind that in the absence of the
quadrupole interaction the distances between the first and second lines
and the fifth and sixth lines in the Mossbauer sextet are equal. 
\begin{table}[!htb]
\caption{\label{spectra-param} The distances between the first and
second ($R_{12}$) and the fifth and sixth ($R_{56}$) lines of the
Mossbauer spectrum in the case of the combined hyperfine interaction
for the positions Fe I and Fe II and for the summary spectra.}
\begin{tabular}{|c|c|c|c|c|c|c|}
\hline
 &\multicolumn{3}{c|}{prismatic}&\multicolumn{3}{c|}{octahedral} \\
\hline
       & Fe I&Fe II&summary& Fe I&Fe II&summary \\
\hline
$R_{12}$& 1.71& 1.43&  1.52 & 1.43& 1.33& 1.38 \\
$R_{56}$& 1.21& 1.52&  1.50 & 1.43& 1.64& 1.57 \\
\hline
\end{tabular}
%\raggedcenter
     \end{table}

In an actual non-equilibrium sample, an even more complicated case
should be realized, as there should be a mixture of the $Fe_3C$
modifications, and only a scrupulous mathematical processing of spectra
with a preliminary model of the relations between the HF parameters
obtained theoretically for two modifications can give a necessary
evidence of the octahedral modification.

The quadrupole splitting above the Curie temperature and its absence 
below have been also observed earlier in experiments of \cite{Ron}. The
parameter of quadrupole splitting measured above $T_C$ in \cite{Ron}
corresponds to our calculation (see \cite{Our-CM}). However, the 
authors supposed that the disappearance of the quadrupole interaction
in the spectrum below $T_C$ is a result of a special angle between EFG
and HFF (55 degree). Our calculations show that this is not true for all
positions of iron atoms and for both modifications (see
Table~\ref{calc-HFparam}). As the MS simulation shows, the presence of a
large quadrupole interaction is concealed under the Curie temperature
by superposition of few subspectra.

Observing the change in the carbon position upon annealing the
plastically-deformed cementite is hardly possible using standard simple
approaches to the MS. A first sign of existence of the octahedral
modification in samples and its disappearance with annealing is a change
of the width of lines in the MS and of the relative distances between
lines (compare the MS's marked as 3 in Fig.~\ref{Moss_bTc}).

So, to construct an initial model of the local atomic structure for a
mathematical processing of MS, a simple approach does not serve: it is
not possible to finally receive an adequate fit of the MS. A correct 
description can be obtained using the results of first-principles
calculations. Certainly, one must bear in mind that first-principles
calculations do not give very accurate magnitude of hyperfine
parameters and give correctly the tendencies in their behavior and the
relations between different sites.

\begin{acknowledgement}
The authors are grateful to professor V. Rusakov for helpful 
discussions. This work was partially supported by RFBR (grant
07-03-96011).
\end{acknowledgement}

\end{document}